\documentclass[10pt,letterpaper]{article}
\usepackage[top=0.85in,left=1.0in,footskip=0.75in,marginparwidth=2in]{geometry}

\usepackage[utf8]{inputenc}

\usepackage{cite}

\usepackage{nameref,hyperref}

\usepackage[right]{lineno}

\usepackage{microtype}
\DisableLigatures[f]{encoding = *, family = * }

\usepackage{changepage}

\usepackage[aboveskip=1pt,labelfont=bf,labelsep=period,singlelinecheck=off]{caption}

\makeatletter
\renewcommand{\@biblabel}[1]{\quad#1.}
\makeatother

\usepackage{lastpage,fancyhdr,graphicx}
\usepackage{epstopdf}
\fancyheadoffset[L]{2.25in}
\fancyfootoffset[L]{2.25in}

\usepackage{color}

\definecolor{Gray}{gray}{.25}

\usepackage{graphicx}

\usepackage{sidecap}

\usepackage{wrapfig}
\usepackage[pscoord]{eso-pic}
\usepackage[fulladjust]{marginnote}
\reversemarginpar

\begin{document}
\vspace*{0.35in}

% title goes here:
\begin{flushleft}
{\Large
\textbf\newline{Growth Mode Selection of Radially Growing Turing Patterns}
}
\newline
% authors go here:
\\
Noah H. Somberg,
Christopher Konow,
Irving R. Epstein,
Milos Dolnik\textsuperscript{*}
\\
\bigskip
\bf Department of Chemistry, Brandeis University, Waltham, MA, USA
\\
\bigskip
* Corresponding author: MS 015 Brandeis University, P.O. Box 549110, Waltham, MA, 02454, USA. Email: dolnik@brandeis.edu

\end{flushleft}

\section{Introduction}
Although best recognized for his work in computer science, Alan Turing also produced a foundational paper on biological pattern formation, “The Chemical Basis of Morphogenesis.”\cite{Turing1952TheMorphogenesis} In this seminal work, Turing set out to propose a mechanism for how a stable, homogeneous state can become unstable and generate patterns. In particular, he was interested in how an initially spherically symmetric bundle of cells can become an asymmetric, non-spherical organism which (much to the disappointment of physicists) includes both cows and humans. The model Turing derived relies upon the diffusion and reaction of chemical species.

Turing patterns, which are stationary in time and periodic in space, appear as characteristic spots, stripes, e.g. hexagonal arrays or labyrinth patterns. Turing patterns are ubiquitous in nature, arising, for example, in leopard spots,\cite{Liu2006Two-stageJaguar} fish skins,\cite{Kondo1995APomacanthus} and desert vegetation.\cite{Rietkerk2004} Turing-type mechanisms have also been implicated in kidney branching,\cite{Menshykau2019} limb development,\cite{Badugu2012} ant burial habits,\cite{Ball2015},\cite{Theraulaz2002} and the distribution of crime in cities.\cite{Ball2015},\cite{Short2010} Turing patterns occur in reaction-diffusion systems of activators and inhibitors, where the inhibitor typically diffuses much faster than the activator.\cite{Murray2003} The development of Turing patterns has also been studied on growing systems.\cite{Klika2017HistoryInstability}

Continuing our previous work\cite{Konow2019}, we study the growth of Turing patterns both experimentally and numerically. We utilize the chlorine dioxide-iodine-malonic acid (CDIMA) reaction to experimentally produce and grow Turing patterns, the formation of which we simulate with the Lengyel-Epstein model. In this work we utilize COMSOL® to validate our previous observations of how Turing pattern morphologies are selected based on growth rate. We also explore additional methods of Turing pattern growth, including exponential growth of Turing patterns and pattern mode selection with partial suppression. 

\section{Theory}
Due to the challenging logistics and time scales associated with the formation of biological Turing patterns, chemical systems provide a more controlled and reproducible medium to study the patterns. One such Turing pattern-generating chemical system is the CDIMA reaction. The reaction produces Turing patterns in a continuously fed unstirred reactor (CFUR) within a gel. The CDIMA reaction is photosensitive; the formation of the Turing pattern is inhibited by the photodissociation of iodine. Bright visible light will suppress the Turing patterns, causing the system to adopt a homogenous state.\cite{Munuzuri1999ControlIllumination} This feature of the CDIMA reaction makes it an ideal candidate to study the growth of a Turing system.  Growth can be simulated by projecting a uniform field of bright light onto the CFUR, inhibiting the reaction, then allowing a shaded region (where the Turing pattern is able to form) to increase in size.

Previous experimental and numerical results by our group have demonstrated that varying the growth rate of the Turing patterns will select different pattern morphologies.\cite{Konow2019},\cite{Miguez2006EffectFormation} These studies utilized a limited, but experimentally accessible model of growth; the size of a dark domain is increased while surrounded by an illuminated region. As the growth proceeds, area is removed from the lit domain and added to the dark domain so that the Turing pattern can form in a larger area. This is performed at varying growth rates, which then induce corresponding pattern morphologies. 

To validate our experimental system and examine the robustness of the observed Turing pattern selection trends, we utilize COMSOL Multiphysics® (versions 4.4, 5.3a, and 5.4) to examine the growth of Turing patterns in a more elementary but experimentally inaccessible system.

\section{Model \& Methods}
To produce Turing patterns we utilize the Lengyel-Epstein (L.E.) two variable model\cite{Lengyel1991ModelingSystem} of the CDIMA reaction modified to account for illumination.\cite{Munuzuri1999ControlIllumination} The L.E. model is given in Equations \ref{eq1} and \ref{eq2}.

\begin{equation} 
    \frac{\partial u}{\partial \tau} = a-u-\frac{4uv}{1+u^2}-W+\nabla^2u
    \label{eq1}
\end{equation}
\begin{equation}
    \frac{\partial v}{\partial \tau} = \sigma \left[b\left(u-\frac{uv}{1+u^2}+W\right)+d\nabla^2v\right] \label{eq2}
\end{equation}

In the model $u$ and $v$ are associated with the dimensionless concentrations of the activator (iodide) and inhibitor species (chlorine dioxide). $\tau$ is dimensionless time, and $a$ and $b$ are dimensionless parameters relating to initial concentrations of reactants. $\sigma$ characterizes the complexation of the activator by an immobile indicator, and $d$ is the ratio of activator and inhibitor diffusion constants. $W$ represents the effect of illumination, where $W = 0$ represents total darkness (no pattern suppression). $W = 1.5$ was used as the maximum light intensity (complete pattern suppression). The L.E. model is simulated numerically on varying growing domains to investigate the behavior of the growth of Turing patterns. The values $a = 12$, $\sigma = 50$, and $d = 1$ were used for all simulations. $b$, which affects the “spottiness” of the pattern, is varied between 0.29 and 0.35.

\subsection{Two-Domain Growth}

As in our previous experimental and numerical investigations,\cite{Konow2019} the Turing patterns are first grown utilizing a two-domain setup, as pictured in Figure 1a. Two domains of concentric circles are defined and assembled with the form union command. The L.E. model is applied to each domain with Coefficient Form PDE physics. Neumann boundary conditions are imposed with zero flux along the exterior of the outer (larger) circle. The outer circle is illuminated ($W_2 = 1.5$) while the interior circle is dark. ($W_1 = 0$). The default extra-fine free-triangular mesh was applied to the geometry.

The pattern growth is first initialized by shading both domains ($W1 = W2 = 0$) and developing the pattern from randomized initial conditions with the time-dependent solver for 1000 time units (t.u.). From this initial pattern growth, the outer domain is again suppressed ($W2 = 1.5$) and the interior domain is set to a small initial size (radius = 0.1). The time-dependent solver is used to develop the pattern for 10 t.u. before increasing the size of the patterned/dark domain with the parametric sweep. This continues until the dark domain reaches a maximum radius of 64 space units (s.u.). The step sizes vary between 0.5 and 7.5 s.u., giving a range of radial growth rates of 0.05 to 0.75 s.u./t.u

\begin{figure}
  \centering
  \includegraphics[height=8cm]{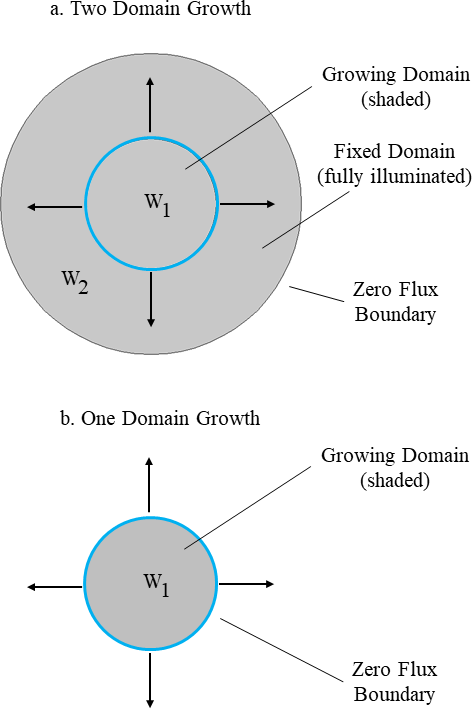}   
  \caption{1a. Simulation scheme two domain growth. 
1b. Simulation scheme for one domain growth.}
  \label{fig:twodomainscheme}
\end{figure}

\subsection{One-Domain Growth}

We investigate single-domain growth both to simplify the simulations and investigate the robustness of the previously observed selection trend. Additionally, since biological growth typically occurs with “one–domain” we utilize several types of one-domain growth to model biological growth more closely. In one-domain growth, the size of a single domain of developing Turing patterns is increased at varying growth rates, without the illumination forcing, as pictured in Figure 1b. Several meshing methods were used in the finite element solver to validate our model and experimental setup. For animations of each meshing method, see Appendix B.

\subsubsection{Addition Type Growth}

In the two-domain growth, as the system expands the finite elements are effectively transferred from the illuminated domain to the dark domain as the system expands; i.e., the dark domain starts with very few elements and gains them as the system size increases. To replicate this process in one-domain growth we fix the size of the mesh elements (maximum element size = minimum element size = 1 s.u.) so that as growth occurs new elements are added to the exterior of the domain, emulating the two-domain growth in a one domain system.

\subsubsection{Stretching Type Growth}

Alternatively, a finer resolution at the initialization of the growth is achieved if the same number of elements are used throughout the simulations, and they increase in size, “stretching” as the domain grows. The default extra-fine free triangular mesh is applied to the domain, which gives the same number of mesh elements throughout the sweep. 

\subsubsection{Growth and Division Type Growth}

A third growth scheme we examined is a compromise between replicating the two-domain growth and achieving fine resolution at small sizes. The mesh is configured to proceed via stretching (elements increase in size) stepwise until the elements reach a critical size where the elements “split” and are refined to smaller elements. These new elements then increase in size as before with the growth, until they “split” again as the mesh is refined. This meshing method is an intermediate between stretching and addition, where the resolution is higher than in addition, but new elements are still added during the growth.

\subsection{Linear and Exponential Growth}

As the growth and division type growth naturally lends itself to period doubling, (i.e., elements increase in size, then split in half at some defined rate) we also investigated the behavior of each of the proceeding four types of growth in both a linearly and exponentially growing system. 

In the linear system we grow the domains by fixed step size between 0.05 s.u./t.u. to 0.75 s.u./t.u., whereas in exponential growth the doubling speed is varied by altering the size of the steps from one element size to another. 0.1 s.u. is slow doubling (ten steps to double) whereas 1.0 s.u. is fast doubling (one step to double).

\subsection{Logistic Illumination and Partial Suppression}

\begin{figure}
  \centering
  \includegraphics[height=8cm]{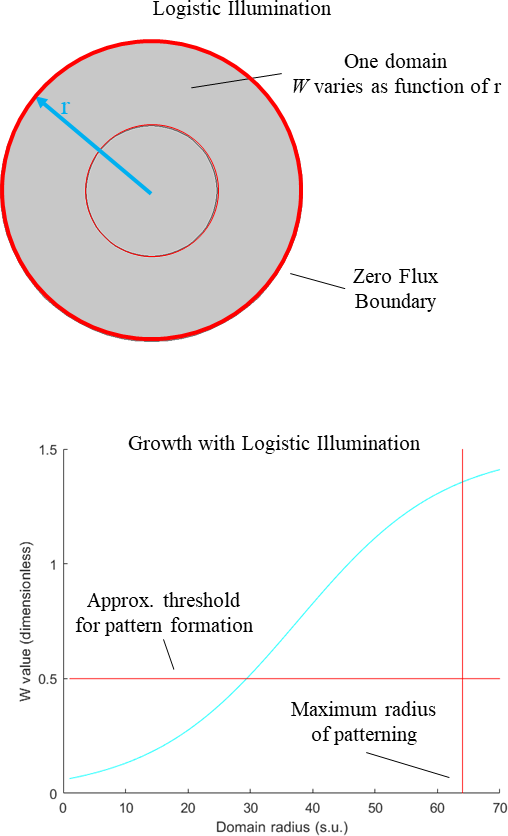}   
  \caption{Simulation scheme for logistic illumination with plot showing illumination curve at one radius size.}
  \label{fig:logisticscheme}
\end{figure}

An alternative growth scheme, which may also be experimentally accessible, is logistic illumination using partial light suppression. Rather than using two domains we apply a radially dependent illumination mask to one domain. We then alter this mask over time to allow for the growth of the region, which is below the threshold of pattern suppression, growing the domain. The scheme for logistic growth is shown in Figure \ref{fig:logisticscheme}.

The mask is applied according to Equation \ref{eq3},
\begin{equation}
    W = \frac{W_{max}}{1+2e^{-k(t)r+2.5}}
 \label{eq3}
\end{equation}

Where r is the position along the radius of the domain and k(t) is a parameter that governs the shape of the intensity profile. The constants in the equation were chosen to match the minimum radius from previous simulations and give the desired shape.

\begin{figure}
  \centering
  \includegraphics[height=6cm]{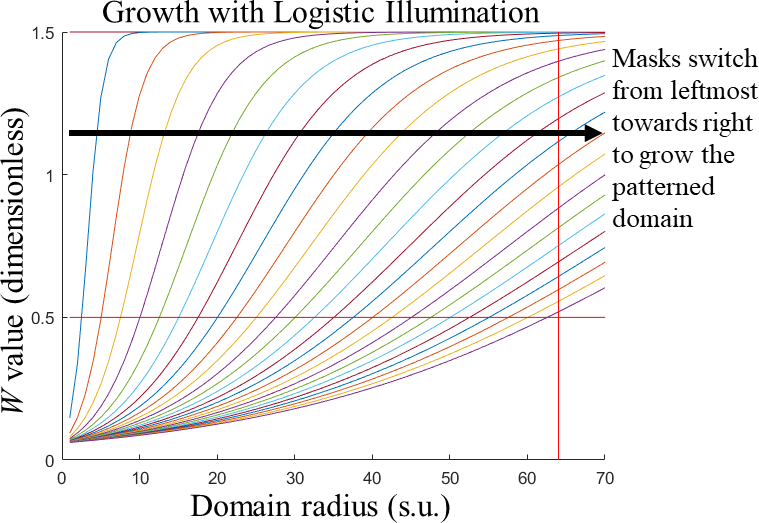}   
  \caption{Varying W masks resulting in domain growth for logistic illumination. The horizontal red line marks the approximate threshold below which patterns can form, and the vertical line marks the desired maximum radius. The mask begins with the furthest left line and proceeds though each mask sequentially towards the right. The lines are calculated to intersect the $W = 0.50$ line at steps of 2.5 s.u., corresponding to a linear growth rate of 0.25 s.u./t.u.}
  \label{fig:logisticplot}
\end{figure}

The value of $k$ is varied with time to give approximately equal steps along the threshold value of $W$ where patterning occurs to match linear growth rates. The parameter k starts very high (500+) giving a very small initial area that can develop Turing patterns. As $k$ decreases to a minimum size of 0.025 the patterns can develop in an increasing area, up to the maximum size of 64 space units. Equation \ref{eq4} was used to calculate k for each step, with r increasing by defined steps. The progression of the logistic masks is shown in Figure \ref{fig:logisticplot}.

\begin{equation}
    k(t) = \frac{2.5 - \ln \left( \frac{W_{max}}{2 W_{treshold}} - \frac{1}{2} \right)}{r}
 \label{eq4}
\end{equation}

Because of the form (Equation \ref{eq3}) chosen for $W$, the interior of the domain quickly falls below the threshold, whereas in the outer parts of the circle the area moves more slowly below the threshold. This scheme simulates growth from the interior of the circle, rather than the exterior as in two-domain growth. This contrasts the previous simulations (with either one or two domains) wherein growth occurred only at the “outside” of the boundary, while no change occurred on the interior.

In previous simulations we have used $W = 0$ for the dark domain and $W = 1.5$ for the illuminated domain. The logistic illumination scheme, however, utilizes a continuum of $W$ values from 0 to 1.5. Therefore, much of the domain is neither fully dark (patterns unsuppressed) nor fully illuminated (patterns completely suppressed). A set of simulations was conducted to observe the effect of just this partial illumination on the Turing pattern morphologies. These were conducted as two-domain simulations, but with varying W values in the dark domain: $W_1$ was varied from 0.125 to 0.750 by increments of 0.125.

\subsection{Simulation Applications}

All four growth schemes were tested in both linear and exponential growth. For each growth scheme eight linear growth rates (0.05 to 0.75 s.u./t.u. by 0.10 increments) and ten doubling rates (0.10 to 1.00 s.u. by 0.10 increments) were tested, and at each growth or doubling rate, the simulation was conducted at seven values of the $b$ parameter (0.29 to 0.35 by increments of 0.01), which influences the natural “spottiness” of the pattern. As this involves over 500 simulations, the COMSOL® Application Builder was used to conduct these efficiently. 

For each type of growth an application was created to perform the initialization and repeat the experiment at a defined $b$ value while sweeping varying growth or doubling rates. The application also exports the resulting growth animations and pictures automatically, as well as notify the user when the simulations are finished so another set can be queued. Each individual simulation at a given set of parameter values took between 15 minutes and 2 hours of computation time on typical desktop machines. Overall this project likely required more than 20 full days of computation time; the application builder ensured that human intervention was minimally required during computation, which allowed for efficient access to the extensive parameter space we investigated. \newline

\subsection{Pattern Wavelength and FFT}
Analysis of certain features of the simulations requires discussion of the Turing pattern wavelength. The pattern wavelength is typically uniform across the domain and so can be obtained via 2D fast Fourier transform (FFT) of the pattern animations across each frame.

\section{Results and Discussion}

In most simulations, four major types of growth are observed. As identified in Konow \textit{et al.} they are inner ring growth (IRG), perpendicular pattern growth (PPG), outer ring addition (ORA), and spotted growth.\cite{Konow2019} 

\begin{figure}
  \centering
  \includegraphics[height=6cm]{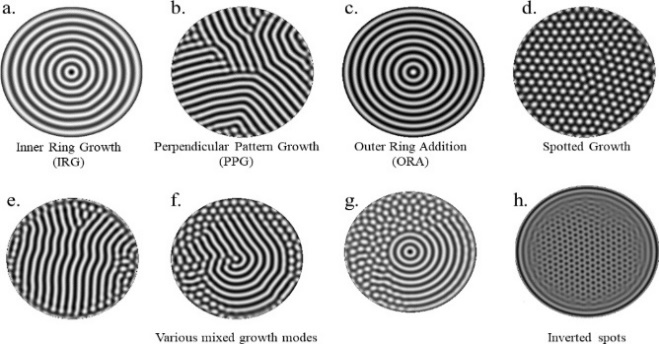}   
  \caption{Examples of observed pattern morphologies. The dark regions correspond to higher $u$ values, representing high triiodide concentration, and the lighter regions correspond to lower $u$ values and therefore lower triiodide concentrations.}
  \label{fig:exmorph}
\end{figure}

Inner ring growth is characterized by the addition of new rings to the center of the domain, as the other rings move outward. An example of the observed final morphology is shown in Figure \ref{fig:exmorph}a.

Perpendicular pattern growth occurs where stripes form so that they are oriented generally orthogonal to the moving boundary. The stripes begin in the center and continue to move outward with the boundary of the Turing pattern. An example of the observed final morphology is shown in Figure \ref{fig:exmorph}b.

Outer ring addition is characterized by the addition of rings of patterns to the exterior of the pattern, while the interior pattern stays relatively constant. Once added there is little movement of each individual ring. The observed final morphology is shown in Figure \ref{fig:exmorph}c. Note the IRG and ORA look very similar in final morphology (concentric circles), but they differ in growth mechanism.

Spotted growth occurs at higher values of $b$. As the spatial orientation of the spots cannot be discerned, it is difficult to garner information about growth modes from these patterns. An example of the spotted growth morphology is shown in Figure \ref{fig:exmorph}d.

Various combinations of the above growth modes are also observed in simulations, although usually with a dominant mode visible. Additionally, in certain experiments “inverted” spots appear, characterized by dark spots surrounded by a light area.  Mixed growth modes are shown in Figure \ref{fig:exmorph}e-g, and inverted spots are shown in Figure \ref{fig:exmorph}h.

\begin{figure}
  \centering
  \includegraphics[height=10cm]{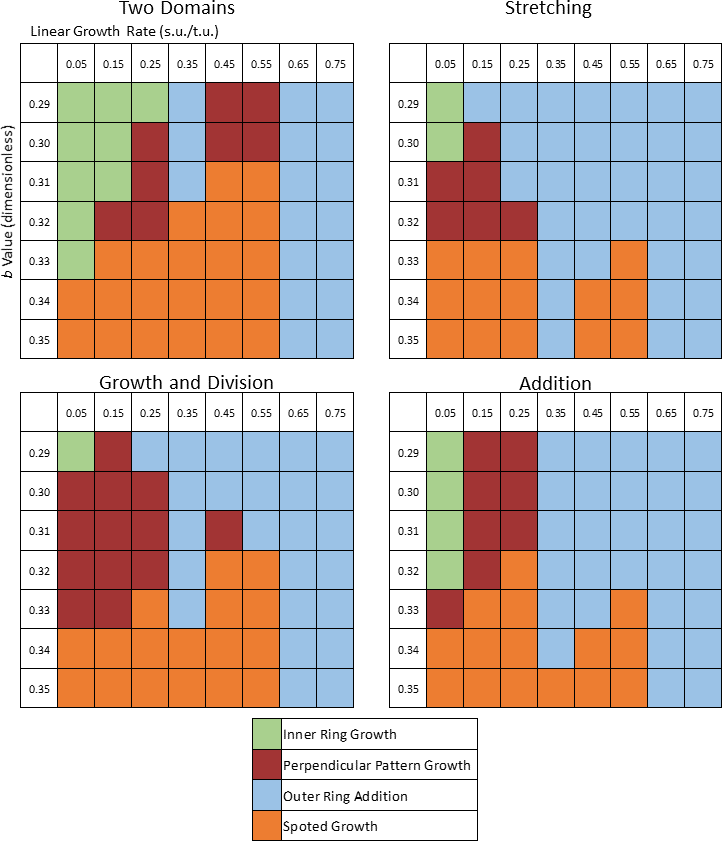}   
  \caption{Observed pattern morphologies for different growth methods at varying linear growth rates and $b$ parameter values. Each square is one simulation assigned to a category of growth.}
  \label{fig:lingrowth}
\end{figure}

\begin{figure}
  \centering
  \includegraphics[height=8cm]{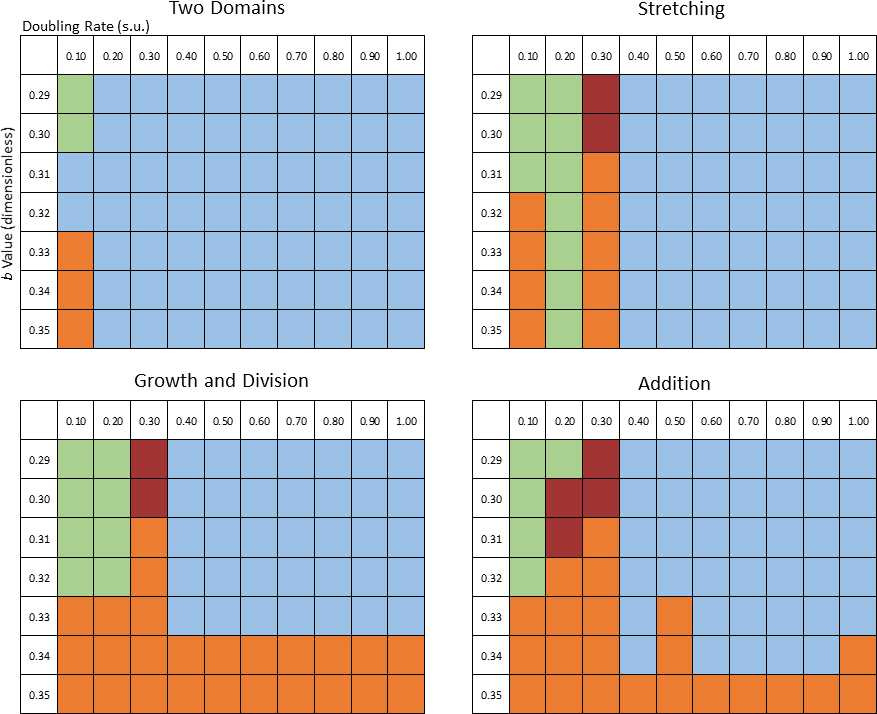}   
  \caption{Observed pattern morphologies for different growth methods at varying exponential growth rates and b parameter values. Each square is one simulation assigned to a category of growth. (See key in Figure \ref{fig:lingrowth})}
  \label{fig:expgrowth}
\end{figure}

\begin{figure}
  \centering
  \includegraphics[height=6cm]{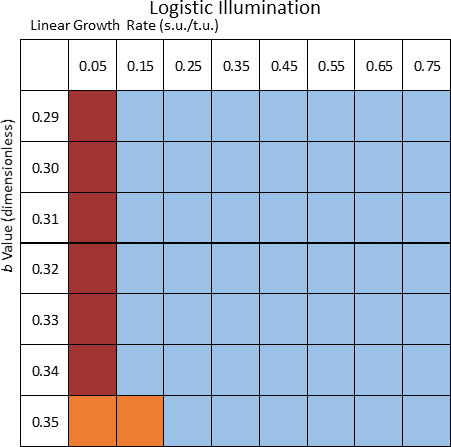}   
  \caption{Observed pattern morphologies for growth with logistic illumination. Each square is one simulation assigned to a category of growth. (See key in Figure \ref{fig:lingrowth})}
  \label{fig:loggrowth}
\end{figure}

\begin{figure}
  \centering
  \includegraphics[height=12cm]{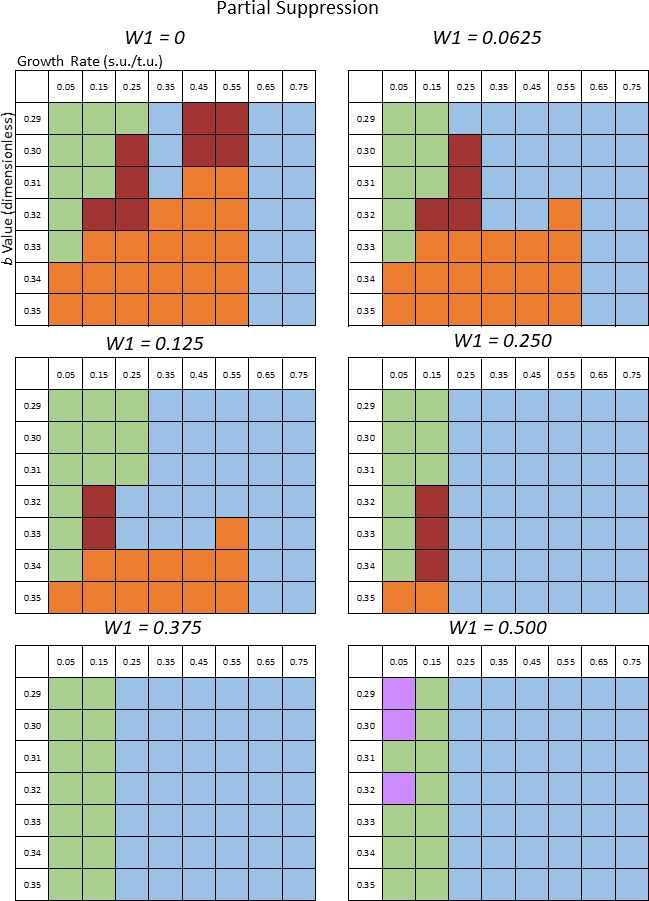}   
  \caption{Observed pattern morphologies for partial illumination simulations. Each square is one simulation assigned to a category of growth. Note $W_1 = 0$ is simply the two domain results from Figure \ref{fig:lingrowth}. (See key in Figure \ref{fig:lingrowth}, with purple representing inverted spots)}
  \label{fig:partgrowth}
\end{figure}

We present simplified charts (figures \ref{fig:lingrowth}, \ref{fig:expgrowth}, \ref{fig:loggrowth}, and \ref{fig:partgrowth}) to demonstrate trends in growth modes, where patterns are represented by colored squares, assigned based on the dominant growth type observed in the animations produced from simulations. Full charts, with all final patterns shown, are included in Appendix A.

\subsection{Linear Growth}

In linearly growing domains, the trend of observed morphologies matches previous observations.\cite{Konow2019} Slow growth favors inner ring growth (IRG). Perpendicular pattern growth (PPG) is observed at intermediate-slow growth rates and may be an intermediary between the fast and slow growth modes. Fast growth favors outer ring addition (ORA). Spotted growth is again observed at high $b$ values at otherwise varying parameters.

Across different growth methods, the trends in the observed morphologies show good agreement with each other, as shown in Figure \ref{fig:lingrowth}. Some local variation is present; two-domain growth seems to favor IRG more than other modes, and PPG appears more in growth and division. However, qualitative trends persist across growth rates and b values. Since in the experimental system the b value may fluctuate through experimental runs, the relative robustness of these pattern selection areas in the parametric space across different meshing methods validates our experimental setup and the accompanying simulations.

A consistent spike of ORA is observed at growth rate 0.35 s.u./t.u. This is observed to some extent at all four growth modes (see Figure \ref{fig:lingrowth}). FFT analysis on a selection of growth speeds and b values indicates that the pattern maintains an intrinsic wavelength near 7.7 s.u. (average for 36 sampled simulations across all four linear growth methods, $s = 0.4$). During the 0.35 s.u./t.u. 3.5 space units are added to the radius of domain at each step, increasing the diameter of the domain by 7.0 s.u. As this is close to the intrinsic wavelength of the pattern, this indicates that the spike in ORA observed near 0.35 s.u./t.u. may be caused by “resonance” between the pattern’s growth rate and the wavelength of the pattern. This is further visible in the animations of the pattern growth (see figures in Appendix B). At the resonant growth rate, the pattern alternates at the boundary, from the boundary being at the center of a dark stripe to the center of a white stripe, adding half a wavelength during each step to the radius, and one full wavelength overall. At higher $b$ values the rings break apart into spots, so the ORA mode is no longer observed.

\subsection{Exponential Growth}

Exponential growth demonstrates similar selection criteria to linear growth, where at slow doubling IRG is observed, and at fast doubling ORA is observed. PPG is observed in a narrow intermediate region, again suggesting it is a transitional pattern between IRG and ORA. As the simulations progressed, they often switched to ORA during the later portion when growth was occurring more quickly. PPG, when present, was often a mixed mode, where growth would occur perpendicular to one boundary and parallel to another, as in Figure \ref{fig:exmorph}e. Exponential growth appears to strongly favor ORA, even at high b values.

The simulations again show good agreement across different growth methods, demonstrating similar trends in observed morphologies, again validating our numerical two-domain model for growth. See the observed morphologies in Figure \ref{fig:expgrowth}.

\subsection{Logistic Illumination and Partial Suppression}

Logistic illumination simulations indicate a different trend, where ORA is heavily favored as the main mode of growth at almost all growth rates, as shown in Figure \ref{fig:loggrowth}. PPG, when present, is typically also a mixed mode, with the pattern forming perpendicular to parts of the boundary and parallel to others. This is a distinctly different trend than what we would expect if logistic illumination corresponded to growth from the interior of the domain; however, the continuum of $W$ values across the domain most likely has a confounding effect.

Partial suppression results indicate that shading above $W = 0$ in the dark domain pushes the system towards ORA as the main growth mode. See figure \ref{fig:partgrowth}. The formation of spots and perpendicular patterns is suppressed, and the system begins to favor ORA as the final morphology. IRG is still visible when present and appears robustly in partial suppression. A new type of growth mode, inverted spots, is also observed in the partial suppression simulations. The combination of these factors indicates that the partial suppression results in an overall shift in parametric space such that previously often observed morphologies, such as spots and perpendicular patterns, are no longer observed.

In the partial suppression results IRG is still present to a fair extent, while it does not appear at all in logistic illumination. As the logistic illumination scheme encourages fast growth from the interior, this may suppress IRG, which is growth which could require the slow increase of the domain interior to occur. The overall tendency of the logistic illumination scheme towards PPG, however, may be a result of the partial suppression effect.

\section{Conclusions}

While local variations in each growth mode are observed, we demonstrate that the pattern trends in the CDIMA simulation are robust both in single and two-domain growth. This validates the experimentally viable two-domain setup. Exponential growth leads to expected growth modes with the fast-growth mode (ORA) being preferred at a larger range of growth rates. The narrow band where PPG is observed indicates that PPG may be a transitional mode between IRG and ORA.

Simulations to replicate “growth from the interior” via logistic illumination do not show simple pattern trends as in one/two-domain systems. It appears that the partial suppression of the illumination across the boundary confounds the previously observed trends, shifting the system in the phase space to a different set of pattern formation mechanisms.

Future investigations could include replicating partial suppression and logistic illumination experimentally and utilizing COMSOL to investigate the growth of Turing patterns in three dimensions.

\section{Supporting Information}
Full versions of parameter space diagrams for each growth type and reaction animations are included in Appendices A and B, respectively. Appendices are available at \url{http://people.brandeis.edu/~nhsomberg/} Viewing the animations may require the document to be opened in Adobe Acrobat Reader. Adobe Flash Player may be required as well.

\section{Acknowledgments}
We thank Václav Klika for his ongoing collaboration and insight into Turing pattern growth. We also acknowledge the financial support of the National Science Foundation (NSF CHE-1856484) and the M.R. Bauer Foundation.

%This is where your bibliography is generated. Make sure that your .bib file is actually called library.bib
\bibliography{comsolpaper}

%This defines the bibliographies style. Search online for a list of available styles.
\bibliographystyle{abbrv}

\end{document}